\documentclass[english,reprint, longbibliography, superscriptaddress, breaklinks=true, showkeys, showpacs=false, nofootinbib]{revtex4-2}

\usepackage[T1]{fontenc}
\usepackage[utf8]{inputenc}
\usepackage{color}
\usepackage{babel}
\usepackage{amsmath}
\usepackage{amssymb}
\usepackage{mathtools}
\usepackage{subfigure}
\usepackage{graphicx}
\usepackage{physics} 
\usepackage{dsfont}
\usepackage{xcolor}
\usepackage{ulem} 
\usepackage{soul} 

\makeatletter
\@ifundefined{textcolor}{}
{%
 \definecolor{BLACK}{gray}{0}
 \definecolor{WHITE}{gray}{1}
 \definecolor{RED}{rgb}{1,0,0}
 \definecolor{GREEN}{rgb}{0,1,0}
 \definecolor{BLUE}{rgb}{0,0,1}
 \definecolor{CYAN}{cmyk}{1,0,0,0}
 \definecolor{MAGENTA}{cmyk}{0,1,0,0}
 \definecolor{YELLOW}{cmyk}{0,0,1,0}
}

\pdfoutput=1
\usepackage[colorlinks=true,allcolors=black]{hyperref}
\usepackage{url}
\usepackage{breakurl}
\makeatother
\usepackage[scr=boondox,  
            ]   
           {mathalpha}

\begin{document}

\title{Thermodynamic entropy production in the dynamical Casimir effect}

\author{Gustavo de Oliveira}
\email{gustav.o.liveira@discente.ufg.br}
\affiliation{QPequi Group, Institute of Physics, Federal University of Goi\'{a}s, 74.690-900, Goi\^{a}nia, Brazil}

\author{Lucas Chibebe C\'{e}leri}
\email{lucas@qpequi.com}
\affiliation{QPequi Group, Institute of Physics, Federal University of Goi\'{a}s, 74.690-900, Goi\^{a}nia, Brazil}

\begin{abstract}
This paper address the question of thermodynamic entropy production in the context of the dynamical Casimir effect. Specifically, we study a scalar quantum field confined within a one-dimensional ideal cavity subject to time-varying boundary conditions dictated by an externally prescribed trajectory of one of the cavity mirrors. The central question is how the thermodynamic entropy of the field evolves over time. Utilizing an effective Hamiltonian approach, we compute the entropy production and reveal that it exhibits scaling behavior concerning the number of particles created in the short-time limit. Furthermore, this approach elucidates the direct connection between this entropy and the emergence of quantum coherence within the mode basis of the field. In addition, by considering a distinct approach based on the time evolution of Gaussian states we examine the long-time limit of entropy production within a single mode of the field. This approach results in establishing a connection between the thermodynamic entropy production in a single field mode and the entanglement between that particular mode and all other modes. Consequently, by employing two distinct approaches, we comprehensively address both the short-term and long-term dynamics of the system. Our results thus link the irreversible dynamics of the field, as measured by entropy production and induced by the dynamical Casimir effect, to two fundamental aspects of quantum mechanics: coherence and entanglement.
\end{abstract}

\maketitle

\section{Introduction}

While the fundamental laws of physics exhibit time-reverse symmetry, we encounter irreversible phenomena in our surroundings when dealing with complex systems. In classical physics, irreversibility is primarily characterized by the second law of thermodynamics, which asserts that the thermodynamic entropy of a closed system cannot decrease over time~\cite{Callen1991}. When fluctuations come into play, stronger principles known as fluctuation theorems emerge~\cite{Esposito2009,Campisi2011}, and irreversible processes are those in which entropy tends to increase on average.

When considering quantum systems, various approaches have emerged in the pursuit of comprehending thermodynamics from a microscopic perspective. Some of these developments include information theory~\cite{Goold2016}, statistical physics~\cite{Strasberg2022}, and axiomatic theories~\cite{Hulse2018}. For a comprehensive exploration of entropy production in both classical and quantum systems, we recommend Ref.~\cite{Landi2021} and its associated references.

We are focusing on the thermodynamics of closed quantum systems, where the time evolution follows a unitary process. This implies that the von Neumann entropy remains constant over time. As a result, this measure is inadequate for quantum thermodynamic entropy because it contradicts the well-established experimental observation that, in general, spontaneous processes tend to increase entropy. Furthermore, it fails to respect the fundamental thermodynamic relation. To tackle this fundamental issue, we turn to the diagonal entropy, as defined in Ref.~\cite{Polkovnikov1} as
\begin{equation}
    S_{d}(\hat{\rho}) = - \sum_{n}p_{n}\ln p_{n},
    \label{eq:diag}
\end{equation}
with $p_{n}$ representing the diagonal elements of the system's density matrix $\hat{\rho}$ in the energy eigenbasis. This quantity has been proposed as the thermodynamic entropy for closed quantum systems since it exhibits several interesting properties, including extensivity, positivity, and the property of vanishing as the temperature approaches zero~\cite{Polkovnikov1}. Furthermore, it possesses a crucial characteristic: it increases for every process, whether unitary or not, that induces transitions in the energy eigenbasis. Only when the system's Hamiltonian changes slowly enough will the diagonal entropy remain unchanged. This aligns with our intuition based on the classical definition of thermodynamic entropy, which does not increase for quasistatic processes~\cite{Polkovnikov2,Polkovnikov3}.

It is worth noting that a closely related quantity known as the observational entropy is defined as a coarse-grained version of the diagonal entropy~\cite{Strasberg2021}. Therefore, the findings presented here also apply within the context of observational entropy.

Information theory have also given rise to a novel approach to thermodynamics, as elucidated by a recent work~\cite{Celeri2023}. In this approach, physical quantities are defined as those invariant under the action of a gauge group, and the emerging concept of entropy precisely aligns with the diagonal entropy discussed above. This alignment resonates with the fact that the gauge-invariant definition of heat is intricately tied to transitions within the energy eigenbasis~\cite{Celeri2023}. This observation also establishes a connection between our findings and another cornerstone of physics, the gauge principle.

We can think about this entropy as a measure of the randomness within the energy eigenbasis. Imagine that we only have access to energy measurements of a quantum system, a common limitation when dealing with systems of a sufficiently large dimension where quantum state tomography becomes impractical~\cite{Medeiros2018}. In a general process, whether unitary or not, transitions between energy levels are induced, leading to the development of quantum coherence and potentially entanglement among different parts of the system. The diagonal entropy quantifies the information loss resulting from our limited set of measurements. We refer the reader to Ref.~\cite{Polkovnikov1} for more details regarding such quantity, including its relation to thermodynamics. The aim of the present work is to apply this concept to a quantum field within the context of the dynamical Casimir effect, and explore the relationship between entropy production and quantum properties such as coherence and entanglement.

Specifically, we consider a quantum scalar field confined within a one-dimensional cavity with mirrors in relative motion, a scenario commonly examined in the context of the dynamical Casimir effect~\cite{Moore,DeWitt,Fulling,Davies}. Under specific conditions, this effect predicts the creation of particles from the vacuum due to the dynamic changes of the boundary conditions imposed by the mirror motion. Over the past five decades, numerous developments have appeared in this field, encompassing the impact of imperfect mirrors~\cite{Jackel,BartonI,BartonII,Haro}, distinct geometries~\cite{Dalvit,Mazzitelli,Celeri2008,Pascoal2009,Naylor}, gravitational field effects~\cite{Celeri2009,Fuentes}, nonlinear interactions~\cite{Trunin2021,Trunin2022,Trunin2023}, and entanglement dynamics~\cite{romualdo,DelGrosso2020}. For a comprehensive overview, interested readers are directed to a recent review~\cite{Dodonov2020}.

However, despite these extensive developments, the irreversible dynamics of the quantum field in this scenario have not been explored, to the best of our knowledge. This work aims to begin addressing this gap by focusing on irreversibility, as measured by the increase in quantum thermodynamic entropy ---the diagonal entropy--- associated with the field's dynamics. In other words, how much entropy is generated in the field due to the nonstationary boundary conditions imposed by the motion of the cavity mirrors? We provide answers to this question through two distinct approaches. Firstly, we employ an effective Hamiltonian theory based on Ref.~\cite{law} to calculate the entropy of the total field within the short-time regime. We demonstrate that the entropy increase is intrinsically tied to the generation of quantum coherence within the system's energy eigenbasis, aligning with the gauge theory developed in Ref.~\cite{Celeri2023}. In the second part of the paper, we adopt a different approach to investigate the long-term field dynamics, allowing us to compute the diagonal entropy for a single mode. Interestingly, this entropy is governed by the entanglement between the selected mode and all other modes. These two distinct approaches enable us to connect the irreversibility of field dynamics with two fundamental quantum features: coherence and entanglement.

\section{\label{dceintro}The dynamical Casimir effect}

Let us consider a one-dimensional ideal cavity whose mirrors are located at positions $x=0$ and $x=L(t)$, with $L(t)$ being an externally prescribed trajectory. Confined in this cavity, we have a massless real scalar field $\phi({x},t)$ satisfying the wave equation
\begin{equation}
    \label{waveq} \left(\partial_t^2-\partial_x^2\right)\phi({x},t)=0.
\end{equation}
Given the ideal nature of the mirrors (perfect reflectors), the boundary conditions imposed on the field take the Dirichlet form 
\begin{equation}
\label{dirchlet}
    \phi(0,t)=\phi(L(t),t)=0.
\end{equation}
The set of complex value solutions $\{\phi_{i}\}$ to Eq.~\eqref{waveq} under the restrictions imposed by the non-stationary boundary conditions~\eqref{dirchlet} spams a linear vector space $\mathscr{S}$ with an invariant bilinear form
\begin{equation}
    \label{innerprod}
    \left(\phi_1,\phi_2\right)=i\int_0^{L(t)} \dd x~\left[\phi_1^{*}\partial_t \phi_2-\phi_2\partial_t \phi_1^{*}\right]
\end{equation}
satisfying all the properties of an inner product except for positive definiteness. This last obstacle hinders the use of Eq.~\eqref{innerprod} for the field's decomposition into orthonormal solutions on $\mathscr{S}$. Nevertheless, we can always choose to that matter, any subspace $\mathscr{S}^+\normalsize\subset\mathscr{S}$, as long as it satisfies the following properties: (\textit{i}) the product~\eqref{innerprod} is positive definite on $\mathscr{S}^+$; (\textit{ii}) $\mathscr{S}=\mathscr{S}^+\oplus\overline{\mathscr{S}^+}$ (with the bar designating the complex conjugated of the space) and (\textit{iii}) for all $f^+\in \mathscr{S}^+$ and $f^-\in\overline{\mathscr{S}^+}$, we have $(f^+,f^-)=0$~\cite{Wald}. 

From the last considerations, if we assume the cavity at the interval $t\leq 0$ to be in a static configuration (with constant mirror position $L(t\leq 0)=L_0$), the classical field can be written as
\begin{equation}
\label{phi0}
    \phi(x,t\leq 0)=\sum_k\left[b_{k}f_{k}^{\text{in}}(x,t)+b_{k}^{*}f_k^{\text{in}*}(x,t)\right],
\end{equation}
where the set $\{f_k^{\text{in}}(x,t)\}$ is an orthonormal basis on $\mathscr{S}^+$ while $\{b_k\}$ is a set of complex coefficients. Since the mirrors are at rest, one can use the time translation symmetry of the wave equation as a natural criterion to select $\mathscr{S}^+$ as the space of solutions that oscillates with purely positive frequencies
\begin{equation}
\label{fkin}
    f_k^{\text{in}}(x,t)=\frac{1}{\sqrt{\pi k}}\sin\left(\omega_k^{\text{in}}x \right)e^{-i\omega_k^{\text{in}} t}, \quad \text{for}\ t \leq 0, 
\end{equation}
where $\omega_k^{\text{in}}=k \pi/L_0$ with $k=\{1,2,\dots\}$.

The quantum description of the field is then obtained by means of the usual field quantization prescription. The coefficients $b_k$ and $b_k^{*}$ are promoted to annihilation and creation operators $\hat{b}_{k}$ and $\hat{b}_{k}^{\dagger}$ satisfying the standard commutation relations
\begin{align}
\label{comrel}\comm{\hat{b}_{k}}{\hat{b}_{j}^{\dagger}}=\delta_{{k}{j}} \ \text{and}\ \Big[\hat{b}_{k},\hat{b}_{j}\Big]=\comm{\hat{b}_{k}^{\dagger}}{\hat{b}_{j}^{\dagger}}=0.
\end{align}
The initial vacuum state $\ket{0;\text{in}}$ is defined as the state annihilated by all $\hat{b}_{k}$, whereas a general particle state can be constructed by the application of the creation operator $\hat{b}_{k}^{\dagger}$ on this vacuum state
\[
\ket{\boldsymbol{n};\text{in}}=\ket{n_{{k}_1},n_{{k}_2},\dots;\text{in}}=\prod_{i}\frac{1}{\sqrt{n_{{k}_i}!}}\left(\hat{b}^{\dagger}_{k_{i}}\right)^{n_{{k}_i}}\ket{0;\text{in}},
\]
with $n_{{k}_i}$ representing the number of particles in the ${k}_i$-th mode.

For $t > 0$, when the mirror starts to move, the quantum field can still be decomposed in terms of the initial operators $\hat{b}_k$ and $\hat{b}_k^{\dagger}$ in the form
\begin{equation}
\label{phit}
    \hat{\phi}(x,t> 0)=\sum_k\left[\hat{b}_{k}f_{k}(x,t)+\hat{b}_{k}^{\dagger}f_k^{*}(x,t)\right],
\end{equation}
as long as the new set of mode functions $\{f_k(x,t)\}$ satisfies the conditions: (i) the wave equation \eqref{waveq}, (ii) the time-dependent boundary condition \eqref{dirchlet}, and (iii) the initial condition $f_k(x,0)=f_k^{\text{in}}(x,0)$. In this regard, we proceed by expanding the mode function in a series with respect to an {\it instantaneous basis} $\left\{\varphi_k(x,t)\right\}$ as
\begin{align}
\label{fk}
    f_k(x,t)=\frac{1}{\sqrt{2\omega_k^{\text{in}}}}\sum_j Q_j^{(k)}(t)\varphi_j(x,t),
\end{align}
where
\begin{align}
\label{varphi}
    \varphi_j(x,t):=\sqrt{\frac{2}{L(t)}}\sin\left[\omega_j(t) x\right] \ \ \text{with}\ \omega_j(t)=\frac{j\pi}{L(t)}.
\end{align}
Moreover the Fourier coefficients $Q_j^{(k)}(t)$ introduced in Eq. \eqref{fk} must satisfy the differential equation\footnote{The set of differential equations~\eqref{Qeq} can be obtained by substituting Eq.~\eqref{fk} into the wave equation~\eqref{waveq} and integrating the resulting expression from $0$ to $L(t)$.}
\begin{align}
\label{Qeq}
\ddot{Q}_j^{(k)}
&+\omega_{j}^2(t)Q_j^{(k)}\\
&=\sum_{l}\left[2\lambda(t)g_{kl}\dot{Q}_l^{(k)}+\dot{\lambda}(t)g_{kl}Q_l^{(k)}-\lambda^2(t) h_{kl}Q_l^{(k)}\right],\nonumber\end{align}
together with the initial conditions
\begin{align}
\label{bcQ}
    Q_j^{(k)}(0)=\delta_{jk}, \qquad \dot{Q}_{j}^{(k)}(0)=-i\omega_k^{\text{in}}\delta_{kj}, 
\end{align}
where the upper dot indicates total time derivative, $\lambda(t)=\dot{L}(t)/L(t)$ and the antisymmetric coefficients $g_{kj}$ and $h_{kj}$ are defined for $j\neq k$ as
\begin{align}
    \label{gkj}
    g_{jk} =(-1)^{j-k}\frac{2kj}{j^2-k^2}, \quad \text{and} \quad  
    h_{jk}=\sum_l g_{jl}g_{kl}.
\end{align}

The first noticeable aspect of the provided description is that the mode expansion~\eqref{fk} fundamentally depends on the choice of the basis functions ${\varphi_k(x,t)}$. This occurs because when the time dependence of the boundary condition~\eqref{dirchlet} is taken into account, the natural criterion of selecting solutions with purely positive frequency is no longer available and there is no unambiguous choice for $\mathscr{S}^+$. Consequently, during the cavity motion, the expansion of the field in terms of creation and annihilation operators becomes arbitrary, implying the nonexistence of a preferred choice for a vacuum state. Thus, unless we can specify a measurement process, the usual notion of particle loses its well-defined meaning, and only when the cavity comes to rest we can associate a definite particle interpretation to the quanta described by these operators~\cite{law}.

If the cavity returns to a static configuration after some interval of time $T$ (with a final constant mirror position $L(t \geq T)=L_T$), one can reintroduce a preferred choice for the mode functions as
\begin{equation}
\label{fkout}
    f_k^{\text{out}}(x,t)=\frac{1}{\sqrt{\pi k}}\sin\left(\omega_k^{\text{out}}x \right)e^{-i\omega_k^{\text{out}} t} \quad \text{for}\ t \geq T 
\end{equation}
with purely positive frequencies $\omega_k^{\text{out}}=k \pi/L_T$. Consequently, the initial operators $\hat{b}_k$ and $\hat{b}_k^{\dagger}$ cease to have a physical significance and the field is now decomposed as
\begin{align}
\label{phif}
    \hat{\phi}(x,t\geq T)=\sum_k \left[\hat{a}_k f_k^{\text{out}}(x,t)+\hat{a}_k^{\dagger} f_k^{\text{out} *}(x,t)\right],
\end{align}
with the set operators $\hat{a}_k$ and $\hat{a}_k^{\dagger}$ satisfying analogous commutation relations as in Eq.~\eqref{comrel} and defining a new vacuum state $\ket{0;\text{out}}$ as the state annihilated by all $\hat{a}_k$.

As pointed out in Ref.~\cite{romualdo}, although both sets $\{f_k^{\text{in}},f_k^{\text{in}*}\}$ and $\{f_k^{\text{out}},f_k^{\text{out}*}\}$ form a basis for the space of solutions $\mathscr{S}$, they represent different decompositions into the subspaces $\mathscr{S}^+$ and $\overline{\mathscr{S}^+}$. The two sets of mode functions \eqref{fkin} and \eqref{fkout} should then be related by a linear transformation
\begin{align}
\label{fkab}
f_k^{\text{in}}=\sum_j \left[\alpha_{jk}f_j^{\text{out}}+\beta_{jk}f_j^{\text{out}*}\right],
\end{align}
where $\alpha_{jk}$ and $\beta_{jk}$ are complex numbers called Bogoliubov coefficients. Inserting Eq.~\eqref{fkab} into the field decomposition~\eqref{phi0}, and comparing with Eq.~\eqref{phif}, we obtain the set of Bogoliubov transformations
\begin{equation}
\label{bogoliubov}
\hat{a}_{j}=\sum_{k}\left[\alpha_{kj}\hat{b}_{k}+\beta_{kj}^*\hat{b}_{k}^{\dagger}\right].
\end{equation}

Observe that the vacuum defined by $\hat{a}_{k}$ and $\hat{b}_{k}$ are not equivalent in general. As a consequence, when computing the number of particles defined by the final operators $\hat{a}_k$ and $\hat{a}_k^{\dagger}$ with respect to the initial vacuum $\ket{0;\rm{in}}$, results
\begin{equation}
\label{Nbogo}
    N=\langle 0;\text{in}|\sum_{j}\hat{a}_{j}^{\dagger}\hat{a}_{j}|0;\text{in}\rangle = \sum_{kj}|\beta_{jk}|^2.
\end{equation}
In general, $\beta_{jk}$ is non-zero when time-dependent boundary conditions are imposed on the field. This last equation characterizes the DCE as the quantum field phenomenon of particle creation from the vacuum due to the time-dependent nature of the imposed boundary conditions.

Our aim here is to study the entropy generated in the field due to this effect. To start, the next section introduces an effective Hamiltonian approach~\cite{razavy,law,plunien,Haro} to describe the field dynamics. This will be important for us to compute the evolved state and, consequently, the entropy generated by the particle creation process. A limitation of this technique is that it only allow us to study the short-time dynamics of the system as it relies on perturbation theory. Nonetheless, it grants us access to the entire state, enabling the exploration of the relationship between irreversibility and the emergence of quantum coherence.

\section{\label{effective}Effective Hamiltonian Approach}

In this section, we introduce an effective Hamiltonian for the DCE following the developments presented in Ref.~\cite{law}. To accomplish this, we begin by expanding the field operator $\hat{\phi}$ and its conjugate momentum $\hat{\pi}=\partial_t \hat{\phi}$ in terms of the instantaneous basis defined in Eq.~\eqref{varphi}
\begin{subequations}
\label{phipi}
    \begin{align}
    \hat{\phi}(x,t)&=\sum_k \hat{q}_k(t) \varphi_k(x,t),\\
    \hat{\pi}(x,t)&=\sum_k \hat{p}_k(t) \varphi_k(x,t),
\end{align}
\end{subequations}
where the operators $\hat{q}_k(t)$ and $\hat{p}_k(t)$ are defined as
\begin{subequations}
\label{qpint}
\begin{align}
\hat{q}_k(t)&:=\int_0^{L(t)}\dd x~\hat{\phi}(x,t)\varphi_k(x,t),\\ \hat{p}_k(t)&:=\int_0^{L(t)}\dd x~\hat{\pi}(x,t)\varphi_k(x,t).    
\end{align}
\end{subequations}

Comparing Eqs.~\eqref{phipi} with the field operator \eqref{phi0} and its time derivative, the expressions for $\hat{q}_k(t)$ and $\hat{p}_k(t)$ can be computed
\begin{subequations}
    \begin{align}
    \hat{q}_k(t\leq 0)&=\frac{1}{\sqrt{2\omega_k^{\text{in}}}}\left[\hat{b}_k e^{-i\omega_k^{\text{in}}t}+\hat{b}_k^{\dagger} e^{i\omega_k^{\text{in}}t}\right],\\
    \hat{p}_k(t\leq 0)&=i\sqrt{\frac{\omega_k^{\text{in}}}{2}}\left[\hat{b}_k^{\dagger} e^{i\omega_k^{\text{in}}t}-\hat{b}_k e^{-i\omega_k^{\text{in}}t}\right].
\end{align}
\end{subequations}

For $t>0$ the cavity is in motion and an effective description of the field dynamics can be obtained by introducing the decomposition~\cite{law}
\begin{subequations}
\label{qpt}
    \begin{align}
    \hat{q}_k(t)&=\frac{1}{\sqrt{2\omega_k(t)}}\left[\hat{a}_k(t) e^{-i\Omega_k(t)}+\hat{a}_k^{\dagger}(t) e^{i\Omega_k(t)}\right],\\
    \hat{p}_k(t)&=i\sqrt{\frac{\omega_k(t)}{2}}\left[\hat{a}_k^{\dagger}(t) e^{i\Omega_k(t)}-\hat{a}_k(t) e^{-i\Omega_k(t)}\right],
\end{align}
\end{subequations}
where $\Omega_k(t)=\int_0^t dt^{\prime}\omega_k(t^{\prime})$ and the {\it instantaneous} annihilation and creation operators $\hat{a}_k(t)$ and $\hat{a}_k^{\dagger}(t)$ satisfy the standard equal times commutation relations
\[\comm{\hat{a}_k(t)}{\hat{a}_k^{\dagger}(t)}=\delta_{kj}; \Bigl[\hat{a}_k(t),\hat{a}_k(t)\Bigr]=\comm{\hat{a}_k^{\dagger}(t)}{\hat{a}_k^{\dagger}(t)}=0.\]

Here, the name instantaneous refers to the physical interpretation that if we freeze the system at some instant $t_0$, the operators $\hat{a}_k(t_0)$ and $\hat{a}_k^{\dagger}(t_0)$ must describe the particle notion for the field as if the cavity mirror had stopped at position $L(t_0)$. One can recognize the initial and final operators to be $\hat{b}_k:=\hat{a}_k(t=0)$ and $\hat{a}_k:=\hat{a}_k(t=T)$.


Taking the time derivative of Eqs.~\eqref{phipi} along with Eqs.~\eqref{qpt} and, after some algebra (see Appendix~\ref{AppendixA} for details), we obtain the following set of differential equations for the annihilation operator
\begin{align}
\label{odeanih}
    {\dot{\hat{a}}_{j}(t)}  =\sum_k\left[A_{kj}(t){\hat{a}_{k}(t)}+B_{kj}^*(t){\hat{a}_{k}^{\dagger}(t)}\right].
\end{align}
The equation for the creation operator is obtained by simply taking the transpose complex conjugate of this last equation. In this equation, we defined the coefficients
\begin{equation}
\label{Hamcoef}
\left.\begin{array}{l}
A_{kj}(t) \\
B_{kj}(t)
\end{array}\right\}=\frac{1}{2}{\left[\mu_{kj}(t)\mp\mu_{jk}(t)\right]}e^{-i\left[\Omega_k(t)\mp\Omega_j(t)\right]}
\end{equation}

with
\begin{equation}
\label{mu}
    {\mu_{kj}(t)}\coloneqq-\left({\sqrt{\frac{j}{k}}g_{jk}}+\frac{1}{2}{\delta_{jk}}\right)\frac{\dot{L}(t)}{L(t)}.
\end{equation}

Identifying Eq.~\eqref{odeanih} as the Heisenberg equation of motion for the annihilation operator, it is straightforward to write down the effective Hamiltonian {in the Schrodinger picture as}\footnote{{Although Hamiltonian~\eqref{Heff} differs from that in Ref.~\cite{law} due to the absence of a term proportional to $\omega_k(t)$, both descriptions are equivalent, since this contribution is contained in the exponential terms in Eq.~\eqref{qpt}.} }
\begin{align}
\label{Heff}
\hat{H}_{{\text{eff}}}(t) =\frac{i}{2}\sum_{{j} {k}} \Bigg[{A}_{kj}(t){\hat{b}_{j}^{\dagger}\hat{b}_{k}}+{B}_{kj}^*(t){\hat{b}_{j}^{\dagger}\hat{b}_{k}^{\dagger}}-\text{h.c.}\Bigg],
\end{align}    
{where "h.c." stands for hermitian conjugate.} 

Here, we can clearly see the existence of two different contributions. The terms containing the coefficients ${B}_{kj}^*$ and ${B}_{kj}$ govern the process of creation and annihilation of pairs of particles, while the ones proportional to ${A}_{kj}^*$ and ${A}_{kj}$ are responsible for scattering of particles between distinct modes.

From this Hamiltonian we can compute the time evolution of any initial density matrix and, therefore, the thermodynamic entropy given in Eq.~\eqref{eq:diag}. This will be done in the sequence.

\subsection{The density operator}

To investigate the entropy production within the proposed scheme, one first needs to obtain an explicit expression for the system's density operator $\hat{\rho}$ after the cavity returns to its stationary configuration. This can be achieved by finding solutions to the dynamical equation
\begin{equation}
    \label{rhodyneqn}
    {\dot{\hat{\rho}}(t)}=-i\left[\hat{H}_{{\text{eff}}}(t),\hat{\rho}(t)\right].
\end{equation}
Conversely, the complex structure of the effective Hamiltonian poses inherent challenges in solving Eq.~\eqref{rhodyneqn}. To overcome this issue, we narrow our focus to a specific category of problems where the equation of motion for the cavity mirror assumes the following form
\begin{align}
\label{Lt} L(t)=L_0\left[1+\epsilon l(t)\right],
\end{align}
{where $l(t)$ is a smooth function of order unity ---as well as its first time derivative---}, while $\epsilon\ll1$ is a small amplitude.

Since the coefficients in Eq.~\eqref{mu} are proportional to $\dot{L}(t)/L(t)$, it is straightforward to see that the Hamiltonian coefficients given in Eqs.~\eqref{Hamcoef} are proportional to $\epsilon$. As a result, the formal solution to Eq.~\eqref{rhodyneqn} up to second order in $\epsilon$ reads
\begin{align}
\label{densityexpansion}\hat{\rho}&(T)=\hat{\rho}(0)-i\int_0^Tdt^{\prime}\left[\hat{H}_{{\text{eff}}}(t^{\prime}),\hat{\rho}(0)\right]\\
&-\int_0^{T}dt^{\prime}\int_0^{t^{\prime}}dt^{\prime\prime}\left[\hat{H}_{{\text{eff}}}(t^{\prime}),\left[\hat{H}_{{\text{eff}}}(t^{\prime\prime}),\hat{\rho}(0)\right]\right].\nonumber
\end{align}

We are interested in the particular case of the initial vacuum state $\hat{\rho}(0)=\ket{0;\text{in}}\bra{\text{in};0}$, since we want to study the thermodynamics of the particle creation process. It is convenient to write the evolved state in terms of the initial operators ${\hat{b}}_k$ and ${\hat{b}}_k^{\dagger}$, which are related to the operators {$\hat{a}_{k}(t)$} and {$\hat{a}_{k}^{\dagger}(t)$} by the {intanstaneous version of the } Bogoliubov coefficients $\alpha_{kj}{(t)}$ and $\beta_{kj}{(t)}$.

By substituting the transformations~\eqref{bogoliubov} into the set of differential equations~\eqref{odeanih}, we obtain a recursive relation for the Bogoliubov coefficients in terms of powers of $\epsilon$. Up to first order, the resulting coefficients are given by
\begin{subequations}
\label{bogopertu}
    \begin{align}
        \alpha_{kj}({t})&=\delta_{kj}+\int_0^{{t}}dt^{\prime}  {A}_{kj}(t^{\prime}),\\
        \beta_{kj}({t})&=\int_0^{{t}}dt^{\prime}~  {B}_{kj}(t^{\prime}),
    \end{align}
\end{subequations}
which implies
\[
{\hat{a}_{k}(t)}={\hat{b}}_k +\sum_j\left(\tilde{\alpha}_{jk}(t){\hat{b}}_j+\beta_{jk}^*(t){\hat{b}}_j^{\dagger}\right),
\]
where $\tilde{\alpha}_{kj}(t)=\int_0^{t}dt^{\prime}  {A}_{kj}(t^{\prime})$. A direct calculation from Eq. \eqref{densityexpansion} leads us to the following expression for the system's density operator up to second order in $\epsilon$
\begin{widetext}
    \begin{align}    
\label{rhot}\hat{\rho}(T)&=\hat{\rho}(0)-\textstyle\frac{1}{2}\displaystyle\sum_{kj}\Bigg\{\beta_{kj}^*\left(\hat{b}_{k}^{\dagger}\hat{b}_{j}^{\dagger}\hat{\rho}(0)\right)-\textstyle\frac{1}{4}\displaystyle\sum_{n m} \Bigg[\beta_{mn}\beta_{kj}^*\left(\hat{b}_{k}^{\dagger}\hat{b}_{j}^{\dagger}\hat{\rho}(0)\hat{b}_{m}\hat{b}_{n}\right)-\beta_{mn}\beta_{kj}^*\left(\hat{b}_{m}\hat{b}_{n}\hat{b}_{k}^{\dagger}\hat{b}_{j}^{\dagger}\hat{\rho}(0)\right)\nonumber\\
&+\beta_{mn}^*\beta_{kj}^*\left(\hat{b}_{m}^{\dagger}\hat{b}_{n}^{\dagger}\hat{b}_{k}^{\dagger}\hat{b}_{j}^{\dagger}\hat{\rho}(0)\right)+2\tilde{\alpha}_{mn}^*\beta_{kj}^*\left(\hat{b}_{m}^{\dagger}\hat{b}_{n}\hat{b}_{k}^{\dagger}\hat{b}_{j}^{\dagger}\hat{\rho}(0)\right)\Bigg]+\text{\normalsize h.c.}\Bigg\}.
\end{align}
\end{widetext}

Considering the initial vacuum state, the number of particles created inside the cavity due to the DCE takes the form
\begin{eqnarray} 
\label{nnm} N(T)&=&\operatorname{Tr}\left\{\sum_{k}\hat{\rho}(T)\hat{b}_{k}^{\dagger}\hat{b}_{k}\right\}=\sum_{{k}{j}} |\beta_{{k}{j}}|^2,
\end{eqnarray}
in agreement with Eq.~\eqref{Nbogo}, thus showing the consistency of our calculations.

We are now ready to discuss the entropy production due to the particle creation process.

\subsection{Entropy production}

As discussed earlier, we consider the diagonal entropy~\cite{Polkovnikov1}
\begin{equation}
\label{diagonalentr}
S_d(\hat{\rho}) = - \sum_{\boldsymbol{n}} \rho_{\text{diag}}^{(\boldsymbol{n})}\ln\rho_{\text{diag}}^{(\boldsymbol{n})},
\end{equation}
as the main figure of merit for characterizing irreversibility. In this equation, $\rho_{\text{diag}}^{(\boldsymbol{n})}=\bra{ \text{in};\boldsymbol{n} }\hat{\rho}\ket{\boldsymbol{n};\text{in}}$ represent the diagonal elements of the system's density operator in the initial energy eigenbasis.

From the expression of the density operator shown in Eq.~\eqref{rhot}, the diagonal entropy can be directly computed, resulting in
\begin{eqnarray}
\label{Sdt}
    S_d(T) &=& -\left[1-\textstyle\frac{1}{2}\displaystyle N(T)\right]\ln\left[1-\textstyle\frac{1}{2}\displaystyle N(T)\right]\nonumber\\
    &-& \sum_{{k}{j}}\textstyle\frac{1}{2}\displaystyle|\beta_{{k}{j}}(T)|^2\ln\textstyle\frac{1}{2}\displaystyle|\beta_{{k}{j}}(T)|^2.
\end{eqnarray}

We first observe that the entropy production depends on the number of particles created inside the cavity. Secondly, we note that this entropy production is exactly equal to the creation of quantum coherence in the energy eigenbasis of the field. To see this, let us consider the relative entropy of coherence~\cite{cramer}
\[
C(\hat{\rho})=S(\hat{\rho}_{d})-S(\hat{\rho}),
\]
which is a measure of the amount of quantum coherence in a given basis. Here $S(\hat{\rho})=-\Tr \hat{\rho}\ln \hat{\rho}$ designates the von Neumman entropy of $\hat{\rho}$ while $\hat{\rho}_{d}$ is the diagonal operator built from the diagonal elements of $\hat{\rho}$ in the selected basis. Since we are interested in the amount of entropy produced during time evolution, we pick up the initial energy eigenbasis to measure coherence. This is fully justified since we are interested in thermodynamics. Under this choice, we directly see that $S(\hat{\rho}_{d}) = S_{d}(\hat{\rho})$. Since our evolution is unitary and the initial state is pure, we have $S(\hat{\rho})=0$, thus implying that 
\begin{equation}
\label{coherence}
C(\hat{\rho})=S_{d}(T).
\end{equation}
Note that, differently from Eq.~\eqref{Sdt}, such a result is a general one, independent of the perturbation theory used here.

This result implies that we will observe irreversibility (positive entropy production) for every process that creates quantum coherence in the energy eigenbasis of the system. Therefore, reversible processes must be those that are performed slowly enough in order to not induce transitions among the energy eigenstates. This result is in agreement with the discussions presented in Refs.~\cite{Polkovnikov1,Polkovnikov2,Polkovnikov3,Celeri2023}, where both entropy production and heat are associated with processes that generate coherence.

In order to illustrate our results, let us consider that the moving mirror performs harmonic oscillations of the form
\begin{equation}
\label{mtrajectory}
    l(t) = \sin (p{\omega_1^{\text{in}}} t),
\end{equation}
where $p$ is an integer, while {$\omega_1^{\text{in}}$} is the first unperturbed field frequency.

For simplicity, we define the small dimensionless time $\tau=\epsilon{\omega_1^{\text{in}}} T/2$ and assume the case in which the mirror returns to its initial position at time $t=T$ after performing a certain number of complete cycles ($p{\omega_1^{\text{in}}} T=2\pi m$ with $m=1,2,\dots$ ). Using Eqs.~\eqref{gkj} and~\eqref{mtrajectory}, we directly obtain
\begin{align}
\label{bkj}
    |\beta_{kj}(\tau)|&=\left\{\begin{array}{cc}
   \sqrt{kj}~ \tau     &\text{if}\ p= k+j  ,\\
    \frac{2\sqrt{kj}\epsilon p}{p^2-(k+j)^2}\sin \left[\frac{2(k+j)\tau}{\epsilon}\right]    &\text{if}\ {p}\neq k+j. 
    \end{array}\right.
\end{align}

By dropping the rapid oscillating terms, the number of particles created takes the form 
\begin{equation}
    \label{RWAN}
    N(\tau)=\frac{1}{6}p(p^2-1)\tau^2,
\end{equation}
in agreement with Ref.~\cite{Dodonov}. Note that the above expression is valid under perturbation theory involving time, and, therefore, it is a good approximation only when $\tau \ll 1$. 

In this case, the diagonal entropy, our focus of interest here, reduces to
\begin{eqnarray}
\label{Sdtp}
S_d(\tau)&=&\frac{1}{2} N(\tau) \Bigg[1-\ln \frac{1}{2}N(\tau)\nonumber\\
&+&\ln\frac{p(p^2-1)}{6}-\frac{6\operatorname{v}(p)}{p(p^2-1)}\Bigg],
\end{eqnarray}
with
\[
\operatorname{v}(p)=\sum_{k=1}^{p-1} (p-k)k\ln (p-k)k. 
\]

\begin{figure}
    \centering
    \includegraphics[scale=0.5]{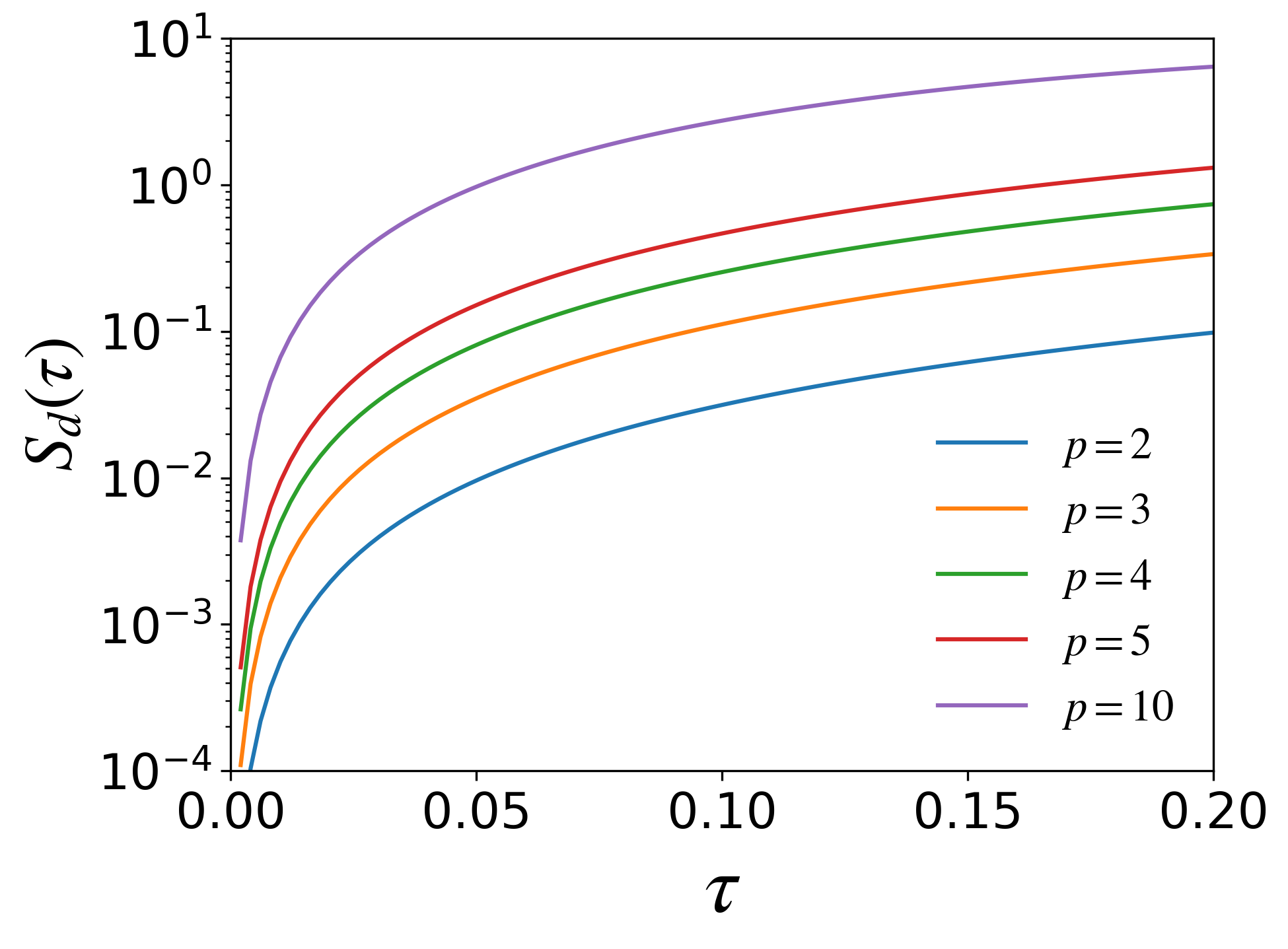}
    \caption{\textbf{Entropy production}. Entropy as a function of $\tau$ for distinct values of the mirror oscillating frequency.}
    \label{fig:N}
\end{figure}

Figure~\ref{fig:N} shows the diagonal entropy for this particular case. As it is clear from the figure, entropy will be produced in the field for every value of the mirror frequency $p$, except for $p=1$, where the number of created particles vanishes.

{The technique employed in this section, based on the effective Hamiltonian, enabled us to calculate the system's entropy production through the time evolution of the density operator. This establishes a direct link between entropy production and the emergence of quantum coherence in the field. Nevertheless, our current analysis is confined to the short-time limit. In the subsequent section, we shift to the Heisenberg picture and quantify entropy production in relation to the time evolution of Gaussian states. This approach permits an exploration of the contribution to entropy production arising from the generation of entanglement between a single mode and the remainder of the field. Therefore, we see that these two approaches are complementary to each other.}

\section{\label{sectionIBD}Gaussian state approach}

The last section presented an analysis of the entropy production constrained to the short-time regime of the entire system. Now, we introduce a different approach that enables us to analyze entropy production in a specific mode across all time intervals. Additionally, this method facilitates the exploration of the entropy dynamics and its connection with the entanglement between the selected mode and all other modes in the system. 

{To achieve this goal we follow the techniques outlined in Ref.~\cite{Dodonov} where the dynamics of the system during the cavity motion is described in the Heisenberg picture. In this approach, the field is decomposed in terms of the Fourier coefficients $Q_j^{(k)}(t)$ through Eq.~\eqref{phit} along with the mode function~\eqref{fk}. Consequently, the dynamics of the system is determined by solving the infinite set of coupled differential equations~\eqref{Qeq} for the Fourier coefficients, with each equation encompassing an infinite number of time-dependent terms.

The problem can be simplified if we consider the special case of parametric resonance, \text{i.e.}, when one of the mirrors undergoes small oscillations at twice the fundamental frequency of the unperturbed field. Therefore,} we impose the following form for the mirror trajectory
\begin{equation}
\label{mirror}
    L(t)=L_0\left[1+\epsilon{\sin} \left(2{\omega_1^{\text{in}}} t\right)\right].
\end{equation}

If the mirror returns to its initial position $L_0$ after some interval of time $T$, then $\omega_k^{\text{in}}=\omega_k^{\text{out}}=\omega_k$ and the right-hand side of Eq.~\eqref{Qeq} vanishes. Under these considerations, it is possible to write
\begin{equation}
\label{bogoliubovQ}
Q_j^{(k)}(t\geq T)=\sqrt{\frac{\omega_k}{\omega_j}}\left(\alpha_{kj}e^{-i\omega_j t}+\beta_{kj}e^{i\omega_jt}\right),
\end{equation}
where $\alpha_{kj}$ and $\beta_{kj}$ are the Bogoliubov coefficients defined in Eq.~\eqref{bogoliubov}.

Since we impose the field to be weakly perturbed by the mirror oscillations \eqref{mirror}, it is natural to search for solutions to $Q_j^{(k)}(t)$ by allowing the Bogoliubov coefficients in Eq.~\eqref{bogoliubovQ} to be functions that vary slowly in time, i.e., $\dot{\alpha}_{kj},\dot{\beta}_{kj}\sim \epsilon$. Then, by substituting Eq.~\eqref{bogoliubovQ} into Eq.~\eqref{Qeq}, {ignoring terms proportional to $\epsilon^2$ (like $\ddot{\alpha}_{kj},\ddot{\beta}_{kj}$ and $\lambda^2$)} and employing the method of slowly varying amplitudes~\cite{Landau}, it is possible to obtain a set of coupled first order differential equations with time independent coefficients in terms of $\alpha_{kj}$ and $\beta_{kj}$. For $k=1$, this set takes the form~\cite{Dodonov1996}
\begin{subequations}
\label{code1}
    \begin{align}
        {\frac{\dd\alpha_{1j}}{\dd \tau}}&=-\sqrt{3}\alpha_{3j}-\beta_{1j} , \\
         {\frac{\dd \beta_{1j}}{\dd \tau}}&=-\alpha_{1j}-\sqrt{3}\beta_{3j} ,
\end{align}
\end{subequations}
whereas for $k> 2$ we obtain
\begin{subequations}
\label{code2}
   \begin{align}
        {\frac{\dd \alpha_{kj}}{\dd \tau}}&=\sqrt{k(k-2)}\alpha_{(k-2),j}-\sqrt{k(k+2)}\alpha_{(k+2),j},  \\
        {\frac{\dd \beta_{kj}}{\dd \tau}}&=\sqrt{k(k-2)}\beta_{(k-2),j}-\sqrt{k(k+2)}\beta_{(k+2),j}.
\end{align}
\end{subequations}
Because of the initial conditions $\alpha_{kj}(0)=\delta_{kj}$ and $\beta_{kj}(0)=0$, all the coefficients with at least one even index vanish.

Complete solutions to the set of equations \eqref{code1} and \eqref{code2} were obtained in Ref.~\cite{Dodonov} in terms of the hypergeometric function. Nonetheless, in this section we will be interested in computing the diagonal entropy generated in particular modes of the field in the regime of parametric resonance~\eqref{mirror}. As a result, for reasons that will become clear later, it will be sufficient to pay attention only to the asymptotic behavior of the Bogoliubov coefficients with the first index equal to 1. 

For $\tau \ll 1$, their expressions read
\begin{subequations}
\label{assymzero}
    \begin{align}
    \alpha_{1(2\mu+1)}&=(\mu+1)K_\mu J_\mu~\tau^{\mu}+\mathcal{O}(\tau^{\mu+2}),\\
    \beta_{1(2\mu+1)}&=-K_\mu J_\mu~\tau^{\mu+1}+\mathcal{O}(\tau^{\mu+3}),
\end{align}
\end{subequations}
with $J_\mu=(2\mu)! / 2^\mu(\mu!)^2$ and $K_\mu=(-1)^{\mu}\sqrt{2\mu+1}/(\mu+1)$, whereas for $\tau \gg 1$
\begin{subequations}
\label{assyminfty}
    \begin{align}
    \alpha_{1(2\mu+1)}&\approx \frac{2}{\pi}\frac{(-1)^\mu}{\sqrt{2\mu+1}},\\
    \beta_{1(2\mu+1)}&\approx \frac{2}{\pi}\frac{(-1)^\mu}{\sqrt{2\mu+1}},
\end{align}
\end{subequations}
with $\mu=0,1,2,\dots$.

Now we are ready to write down the reduced density operator for the considered mode and to address the question of the dynamics of the entropy production and its relation to entanglement. 

\subsection{Reduced density operator}

The reduced density operator of mode $m$ is given by 
\begin{equation}
    \label{tracedrho}
    \hat{\rho}_{m}=\operatorname{Tr}_{\{k\}/m}\hat{\rho},
\end{equation}
where $\operatorname{Tr}_{\{k\}/m}$ denotes the trace of the total density operator $\hat{\rho}$ over all the modes except the $m$-th one.

Now, from the previous section, we can see that the time evolution of the field can be described by an effective quadratic time-dependent Hamiltonian. We know that the time evolution governed by quadratic Hamiltonians transforms any Gaussian state into another Gaussian state, which are completely characterized by the covariance matrix. 

As the vacuum state belongs to the class of Gaussian states, it is in fact possible to describe our initial state in terms of the Wigner function for the $m$-th mode, which reads
\begin{equation*}
W_m(\mathbf{q})=\frac{1}{\sqrt{2\pi\operatorname{det}\mathbf{\Sigma}_{m}}}e^{-\frac{1}{2}\left(\mathbf{q}-\langle\mathbf{q}\rangle\right)\mathbf{\Sigma}_m^{-1}\left(\mathbf{q}-\langle\mathbf{q}\rangle\right)},
\end{equation*}
where $\mathbf{q}=\left(\hat{q}_{m},\hat{p}_{m}\right)$ is the quadrature operator with components
\begin{subequations}
\label{quadrature}
    \begin{align}
    \hat{q}_m&=\frac{1}{\sqrt{2}}\left(\hat{a}_m^{\dagger}+\hat{a}_m\right),\\
    \hat{p}_m&=\frac{i}{\sqrt{2}}\left(\hat{a}_m^{\dagger}-\hat{a}_m\right).
\end{align}
\end{subequations} 
$\mathbf{\Sigma}_{m}$ stands for the covariance matrix
\begin{equation}
\label{covariance}
\mathbf{\Sigma}_{m}\equiv \left(\begin{matrix}
\sigma_m^q & \sigma_m^{qp}\\
\sigma_m^{qp} & \sigma_m^{p}
\end{matrix}\right)
\end{equation}
with
\begin{subequations}
\label{variances}
    \begin{align}
       \label{v1} \sigma_m^q&=\langle\hat{q}_{m}^2 \rangle-\langle\hat{q}_m\rangle^2,\\ 
       \label{v2}\sigma_m^{p}&=\langle\hat{p}_m^2 \rangle-\langle\hat{p}_m\rangle^2,\\
 \label{v3}\sigma_m^{qp}&=\frac{1}{2}\langle\hat{p}_m\hat{q}_m+\hat{q}_m\hat{p}_m \rangle-\langle\hat{q}_m\rangle\langle\hat{p}_m\rangle.
    \end{align}
\end{subequations}
 
Since we are interested in the diagonal entropy, we focus on the diagonal components of the density operator in the energy eigenbasis. For the special case of an initially vacuum state $\ket{0;\textit{in}}$, these diagonal terms can be written as functions of the covariance matrix elements~\cite{Dodonov}
 \begin{align}
 \label{rhom}
 \rho_m^{(n)}&=\frac{2\left[\left(2\sigma_m^q-1\right)\left(2\sigma_m^{p}-1\right)\right]^{n/2}}{\left[\left(2\sigma_m^q+1\right)\left(2\sigma_m^{p}+1\right)\right]^{(n+1)/2}} \nonumber\\
&\times \operatorname{P}_n\left(\frac{4\sigma_m^q\sigma_m^{p}-1}{\sqrt{(4(\sigma_m^q)^2-1)(4(\sigma_m^{p})^2-1)}}\right),
 \end{align}
where $\operatorname{P}_n$ is the Legendre polynomial of order $n$ and $\rho_m^{(n)}=\bra{\text{in};n}\hat{\rho}_{m}\ket{n;\text{in}}$ is the $n$-th diagonal element of the reduced density operator in the initial energy eigenbasis. 

By expressing the quadrature operators \eqref{quadrature} in terms of the initial operators $\hat{b}_{k}$ and $\hat{b}_k^{\dagger}$ defined in Eq.~\eqref{bogoliubov}, the variances can be directly computed, resulting in
\begin{subequations}
\label{qpvar}
    \begin{align}
    \sigma_m^{q}&=\frac{1}{2}\sum_k\left|\alpha_{km}{-}\beta_{km}\right|^2,\\
    \sigma_m^{p}&=\frac{1}{2}\sum_k\left|\alpha_{km}{+}\beta_{km}\right|^2
\end{align}
\end{subequations}
where $m$ is an odd integer and the cross term $\sigma_m^{qp}$ is identically zero for our choice of the initial state.

By taking the time derivatives of these last equations and inserting the recursive relations \eqref{code1} and \eqref{code2}, one can show that
\begin{subequations}
\label{pqab}
    \begin{align}
{\frac{\dd \sigma^{q}_{m}}{\dd \tau}}&=-\left[\alpha_{1m} {-} \beta_{1m}\right]^2 \\
{\frac{\dd \sigma^{p}_{m}}{\dd \tau}}&= +\left[\alpha_{1m} {+} \beta_{1m}\right]^2,    
\end{align} 
\end{subequations}
which depends only on the Bogoliubov coefficients, with the first index equal to 1 (as we have pointed out in the beginning of the section). Moreover, because the definitions \eqref{quadrature}, the differential equations \eqref{pqab} need to satisfy the initial conditions $\sigma_m^q(0)=\sigma_m^p(0)=1/2$.

We now analyze the solutions to these equations in two distinct regimes, the short-time and the long-time. 

\subsection{Short-time regime}

The short time limit is defined by $\tau\ll1$. Inserting Eqs.~\eqref{assymzero} into Eqs.~\eqref{pqab} and integrating over $\tau$, we obtain
\begin{equation}
\label{perturbation}
\left.\begin{array}{l}
\sigma_{2\mu+1}^{q} \\
\sigma_{2\mu+1}^{p}
\end{array}\right\}=\frac{1}{2}\mp \tau^{2\mu+1}J_\mu^2 \left[1\mp K_\mu^2\tau+\mathcal{O}(\tau^2)\right],\nonumber
\end{equation}
with $J_\mu$ and $K_\mu$ defined in Eq.~\eqref{assymzero}.

Plugging Eqs.~\eqref{perturbation} into Eq.~\eqref{rhom} leads to the following expression for the diagonal components of the reduced density operator
\begin{align}
    \label{rholowlim}\rho_{2\mu+1}^{(n)}&=(-1)^ni^{n}J_\mu^n\tau^{n(2\mu+1)}\left(1-K_\mu^4\tau^2\right)^{n/2}\\
    &\times\Bigg[1-(n+1)J_\mu^2\tau^{2\mu+2}\left(K_\mu^2-\frac{1}{2}J_\mu^2\tau^{2\mu}\right)\Bigg]\nonumber\\
    &\times P_n\left[i\tau\left(K_\mu^2-J_\mu^2\tau^{2\mu}\right)\right]+\mathcal{O}(\tau^{2\mu+3})\nonumber.
\end{align}

This expression is what we need to compute the diagonal entropy the $(2\mu+1)$-th mode. Up to the second order in $\tau$ we obtain, for $\mu=0$, the following result
\begin{align*}
S_{d}^1(\tau\ll 1)=\frac{1}{2}N_1(\tau)\bigg[1-\ln\frac{1}{2}N_1(\tau)\bigg],
\end{align*}
while for any other value of $\mu$, we have
\begin{equation*}
    S_d^{2\mu+1}(\tau\ll 1)=N_{2\mu+1}(\tau)\bigg[1-\ln N_{2\mu+1}(\tau)\bigg]+\mathcal{O}(\tau^{2\mu+3}),
\end{equation*}
where $N_{2\mu+1}(\tau)=K_\mu^2J_\mu^2\tau^{2\mu+2}+\mathcal{O}(\tau^{2\mu+3})$ is the number of particles created in the corresponding mode.

Hence, at short-times, the entropy for each mode increases with the number of created particles, aligning completely with the findings outlined in the preceding section. As expected, the current methodology enables an exploration of the long-time dynamics of the entropy production, and we delve into such an analysis in the subsequent discussion.

\subsection{Long-time regime}

The long-time limit is defined by $\tau\gg1$. In this case, by substituting Eqs.~\eqref{assyminfty} into Eqs.~\eqref{pqab}, we obtain the time derivatives of the system's quadrature variances as 
\begin{subequations}
\label{dotqp}
    \begin{align}
\label{dotq}&{\frac{\dd }{\dd \tau}}\sigma_{2\mu+1}^{q}\approx0\\
\label{dotp}&{\frac{\dd }{\dd \tau}}\sigma_{2\mu+1}^{p}\approx\frac{16}{\pi^2(2\mu+1)}.
\end{align}
\end{subequations}
The specific integration constant for Eqs.~\eqref{dotq} varies for each mode and depends on the complete form of the Bogoliubov coefficients~\cite{Dodonov}, but the general behavior is the same: both quadrature variances start with the same value $1/2$ at $t=0$ and end up assuming distinct asymptotic behavior at $\tau\gg 1$, with $\sigma_m^q$ decreasing to a constant value, whereas $\sigma_m^p$ increases almost linearly in time. 

It is now straightforward to compute the single-mode reduced density matrix as 
\begin{align}
\label{rhotmaior}
\rho_m^{(n)}(\tau\gg1)&=C^{(n)}_m \ [\det\mathbf{\Sigma}_m(\tau)]^{-1/2}+\mathcal{O}(1/\tau)                     
\end{align}
where
\begin{equation}
\label{cmn}
C_m^{(n)} = \frac{1}{\sqrt{1+T_m}}\left(\frac{1-T_m}{\sqrt{1-T_m^2}}\right)^n \operatorname{P}_n\left(\frac{1}{\sqrt{1-T_m^2}}\right)
\end{equation}
is a positive real coefficient with $T_m=1/2\sigma^q_{m}$. 

From the above expressions, we can compute the diagonal entropy associated with the $m$-th field mode as
\begin{equation}
\label{eq:entro}
S_{d}^m(\tau\gg1)\approx S_R^m(\tau)+[\det\mathbf{\Sigma}_m(\tau)]^{-1/2}\mathcal{S}_m,
\end{equation}
where $\mathcal{S}_m=-\sum_n C_m^{(n)}\ln C_m^{(n)}$ and $S_R^m(\tau)=\frac{1}{2}\ln \det\mathbf{\Sigma}_m(\tau)$ is the Rényi-2 entropy of the $m$-th mode~\cite{adesso}. It can be shown that the second term in Eq.~\eqref{eq:entro} diverges logarithmically with the system dimension $\mathcal{N}$. This last fact is expected since we are considering a field theory and the number of degrees of freedom of the system is infinite. Moreover, we must remember that entropy is defined up to a multiplicative and an additive constant. So, this last term is not fundamental for the dynamical behavior of entropy.

For the resonant mode $m=1$, we obtain $\sigma^q_1\to2/\pi^2$~\cite{Dodonov}  and $ \sigma^p_1\to16\tau/\pi^2$, leading to the Réniy-2 entropy
\[
S_{R}^1(\tau)\approx\frac{1}{2}\ln \frac{32}{\pi^4}\tau,
\]
which is in agreement with Ref.~\cite{romualdo}\footnote{Here, the argument in the Réniy-2 entropy differs from Ref.~\cite{romualdo} by a factor of $4$. This occurs because the variances defined in the last reference are twice as large as the ones in Eq.~\eqref{variances}.}. In the case of the subsequent mode $m=3$, now $\sigma^q_3\to38/9\pi^2$  and $ \sigma^p_3\to16\tau/3\pi^2$, so we obtain
\[
S_{R}^3(\tau)\approx\frac{1}{2}\ln \frac{608}{27\pi^4}\tau.
\]

Now, since the global state of the field is pure ---initial pure state under unitary evolution---, $S_{R}^m(\tau)$ quantifies the amount of entanglement between the $m$-th mode and all the remaining ones. Therefore, what Eq.~\eqref{eq:entro} is saying to us is that the asymptotic behavior of the diagonal entropy is fundamentally determined by the generation of entanglement between the considered mode and all the others. 

\section{Conclusions}

{This article considers the problem of thermodynamic entropy production within the framework of the dynamical Casimir effect, exploring two distinct approaches. The initial approach, employing an effective Hamiltonian description of field dynamics, provides a connections between entropy production and the generation of quantum coherence in the field's mode basis in the short-time limit. The second approach, which relies on the reduced density operator of an individual mode and it is valid for all times, establishes a connection between entropy growth and entanglement generation between the selected mode and all the others.} 

{Although both approaches can only be compared in the short-time regime, where both predicts that entropy increases with the number of created particles, they provide different, but complementary information about the dynamics of the entropy production due to the dynamical Casimir effect.}

{In summary, the production of thermodynamic entropy in the field due to the dynamical Casimir effect is governed by the generation of quantum coherence in the field's mode basis and entanglement between the modes. Since our initial state is stationary (vacuum), the diagonal entropy cannot decrease~\cite{Polkovnikov1} and, therefore, neither coherence or entanglement.}

{These results can be understood as follows. A coupling between all the field modes arises due to the non-trivial boundary conditions imposed on the field by the motion of the mirror. Such interaction induces transitions among the modes, which lies at the root of the generation of quantum coherence and quantum entanglement. Although the evolution is unitary, irreversibility, which is characterized by entropy production, also arises due to these transitions, as discussed in Refs.~\cite{Celeri2023,Polkovnikov1,Polkovnikov2,Polkovnikov3}. Reversible processes are defined in the limit where the motion is so slow that there is no particle creation, no scattering and, thus, no entropy production. Note that in the considered context, in which we have a resonant cavity trapping the field, there are motions for which no particles will be created and, thus, no entropy will be produced. This is a point that deserves a deeper investigation.}

{Our research enhances the comprehension of the thermodynamics of quantum fields within non-trivial boundary conditions and exploring the impact of quantum coherence and entanglement on such phenomena. Despite this, numerous questions remain unanswered.}

An interesting question that directly emerges concerns the split of the energy into work and heat, where the latter is associated with the irreversible aspect of the process, while the former should be related to the energy that can be extracted from the field after the process~\cite{Francica2020,Plastina2014}. Another related issue involves the statistical description of the field in terms of stochastic entropy production and the fluctuation theorems~\cite{Santos2019}. Furthermore, what role do multiple quantum coherence and multipartite entanglement play in entropy production? How do the thermalization properties of field dynamics frame into this? Lastly, a question arises regarding whether heat and work adhere appropriately to the equivalence principle~\cite{Basso2023}. These are some of the pertinent questions that will be the focus of future investigations.

\begin{acknowledgments}
This work was supported by the National Institute for the Science and Technology of Quantum Information (INCT-IQ), Grant No.~465469/2014-0, by the National Council for Scientific and Technological Development (CNPq), Grants No~308065/2022-0, and by Coordination of Superior Level Staff Improvement (CAPES).
\end{acknowledgments}

\appendix

\section{Derivation of the effective Hamiltonian\label{AppendixA}}
\subsection{Dynamical equations for the instantaneous creation and annihilation operators}
From Eq. \eqref{waveq}, the dynamical equation of motion for a quantum scalar field and its conjugated momentum, can be written as
\begin{subequations}
\label{dyneqn}
\begin{align}
     \partial_t\hat{\phi}(x,t)&=\hat{\pi}(x,t)\\
    \partial_t\hat{\pi}(x,t)&={\partial_x^2} \hat{\phi}(x,t).
\end{align}
\end{subequations}
{By combining Eqs. \eqref{phipi} and \eqref{qpt}, one can express the fields $\hat{\phi}$ and $\hat{\pi}$ and their correspondent time derivatives as}
\begin{subequations}
\label{alleqn}
\begin{align}
        \hat{\phi}&=\sum_{k} \frac{1}{\sqrt{2\omega_{k}}}\left({\hat{a}_ke^{-i\Omega_k}+\hat{a}_k^{\dagger}e^{i\Omega_k}}\right)\varphi_{k},\\
        \hat{\pi}&=i\sum_{k} \sqrt{\frac{\omega_{k}}{2}}\left({\hat{a}_k^{\dagger}e^{i\Omega_k}-\hat{a}_ke^{-i\Omega_k}}\right)\varphi_{k},\\
        \label{dotphi}{\partial_t}\hat{\phi}&=\sum_{k} \frac{1}{\sqrt{2\omega_{k}}}\left({\hat{a}_ke^{-i\Omega_k}+\hat{a}_k^{\dagger}e^{i\Omega_k}}\right)\left({\partial_t}\varphi_{k}-\frac{\dot{\omega}_k}{2\omega_{k}}\varphi_{k}\right)\nonumber\\
        &+\sum_{k} \frac{1}{\sqrt{2\omega_{k}}}\left({\dot{\hat{a}}_ke^{-i\Omega_k}+\dot{\hat{a}}_k^{\dagger}e^{i\Omega_k}}\right)\varphi_{k}+\hat{\pi},\\
        \label{dotpi}{\partial_t}\hat{\pi}&=i\sum_{k} \sqrt{\frac{\omega_{k}}{2}}\left({\hat{a}_k^{\dagger}e^{i\Omega_k}-\hat{a}_ke^{-i\Omega_k}}\right)\left({\partial_t}\varphi_{k}+\frac{\dot{\omega}_k}{2\omega_{k}}\varphi_{k}\right)\nonumber\\
        &+i\sum_{k} \sqrt{\frac{\omega_{k}}{2}}\left({\dot{\hat{a}}_k^{\dagger}e^{i\Omega_k}-\dot{\hat{a}}_ke^{-i\Omega_k}}\right)\varphi_{k}+{\partial_x^2}\hat{\phi},
\end{align}
\end{subequations}
where, for conciseness, we have suppressed the notation of time and spatial dependence in all terms in \eqref{alleqn}. 
Comparing \eqref{dyneqn} with \eqref{dotphi} and \eqref{dotpi}, we can isolate the time derivative of the operators {$\hat{a}_k$ and $\hat{a}_k^{\dagger}$} by computing
\begin{align}
    &\int_0^{L} \dd x\varphi_j\left({\partial_t}\hat{\phi}-\hat{\pi}\right) =\sum_{k}  \frac{1}{\sqrt{2\omega_{k}}}\left({\dot{\hat{a}}_ke^{-i\Omega_k}+\dot{\hat{a}}_k^{\dagger}e^{i\Omega_k}}\right)\delta_{kj}\nonumber\\
   \label{int1} &-\sum_{k} \frac{1}{\sqrt{2\omega_{k}}}\left({\hat{a}_ke^{-i\Omega_k}+\hat{a}_k^{\dagger}e^{i\Omega_k}}\right)\left(G_{kj}+\frac{\dot{\omega}_k}{2\omega_{k}}\delta_{kj}\right)=0
\end{align}
and
\begin{align}
    \label{int2}&\int_0^{L} \dd x\varphi_j\left({\partial_t}\hat{\pi}-\partial_x^2\hat{\phi}\right)\\
    &=i\sum_{k}\sqrt{\frac{\omega_{k}}{2}}\left({\dot{\hat{a}}_k^{\dagger}e^{i\Omega_k}-\dot{\hat{a}}_ke^{-i\Omega_k}}\right)\delta_{kj}\nonumber\\
    &+i\sum_{k}  \sqrt{\frac{\omega_{k}}{2}}\left({\hat{a}_k^{\dagger}e^{i\Omega_k}-\hat{a}_ke^{-i\Omega_k}}\right)\left(G_{jk}+\frac{\dot{\omega}_j}{2\omega_{j}}\delta_{kj}\right)=0\nonumber,
\end{align}
where it was used $\int_0^L dx \varphi_k\varphi_j=\delta_{{kj}}$  and $G_{kj}\coloneqq\int_0^{L}\varphi_k{\partial_t}\varphi_j$. By defining $\mu_{kj}=\sqrt{\frac{\omega_j}{\omega_k}}\left(G_{kj}+\frac{\dot{\omega}_{k}}{2\omega_{k}}\delta_{kj}\right)$ we obtain from \eqref{int1} and \eqref{int2} {the following equations}
\begin{subequations}
\label{dadad}
    \begin{align}
    {\dot{\hat{a}}_je^{-i\Omega_j}+\dot{\hat{a}}_j^{\dagger}e^{i\Omega_j}}&=\sum_{k} \mu_{kj}\left({\hat{a}_je^{-i\Omega_j}+\hat{a}_j^{\dagger}e^{i\Omega_j}}\right),\\
    {\dot{\hat{a}}_je^{-i\Omega_j}-\dot{\hat{a}}_j^{\dagger}e^{i\Omega_j}}&=\sum_{k}  \mu_{jk}\left({\hat{a}_k^{\dagger}e^{i\Omega_k}-\hat{a}_ke^{-i\Omega_k}}\right).
\end{align}
\end{subequations}
From the last system, it is easy to isolate ${\dot{\hat{a}}_{j}(t)}$ and $ {\dot{\hat{a}}_{j}^{\dagger}(t)}$ as
\begin{subequations}
    \label{aaddynmeqn}
    \begin{align}
   \label{da} \dot{\hat{a}}_j{(t)}&=\sum_k\left[A_{kj}{(t)}a_k{(t)}+B_{kj}^*{(t)}a_k^{\dagger}{(t)}\right],\\
   \label{dad} \dot{\hat{a}}_j^{\dagger}{(t)}&=\sum_k\left[A_{kj}^*{(t)}a_k^{\dagger}{(t)}+B_{kj}{(t)}a_k{(t)}\right],
\end{align}
\end{subequations}
with
\begin{subequations}
\label{dynaada}
    \begin{align}
    A_{kj}{(t)}&=\frac{1}{2}\left[\mu_{kj}{(t)}-\mu_{jk}{(t)}\right]e^{-i\left[\Omega_k{(t)}-\Omega_j{(t)}\right]},\\
    B_{kj}{(t)}&=\frac{1}{2}\left[\mu_{kj}{(t)}+\mu_{jk}{(t)}\right]e^{-i\left[\Omega_k{(t)}+\Omega_j{(t)}\right]}.
\end{align}
\end{subequations}
Since $\omega_{k}{(t)}=k\pi/L{(t)}$ and using the definition \eqref{varphi} we can calculate
\begin{align}
    G_{kj}{(t)}&=g_{kj}\frac{\dot{L}{(t)}}{L{(t)}},\\
    \frac{\dot{\omega}_k{(t)}}{\omega_k{(t)}}&=-\frac{\dot{L}{(t)}}{L{(t)}},
\end{align}
where $g_{kj}$ has the same form as expressed in \eqref{gkj}. So we obtain
$\mu_{kj}{(t)}=-\left(\sqrt{\frac{j}{k}}g_{jk}+\frac{1}{2}\delta_{kj}\right)\frac{\dot{L}{(t)}}{L{(t)}}$ as in Eq. \eqref{mu}.
\subsection{Effective Hamiltonian}

To find the effective Hamiltonian that generates the dynamical equations \eqref{dynaada} we begin by considering the most general quadratic operator
\begin{align}
    \label{ht}\hat{H}{(t)}=\sum_{kl}&\left[\mathcal{A}_{kl}{(t)}\hat{a}_k^{\dagger}{(t)}\hat{a}_l^{\dagger}{(t)}+\mathcal{B}_{kl}{(t)}\hat{a}_k^{\dagger}{(t)}\hat{a}_l{(t)}\right.\\
    &\left.+\mathcal{C}_{kl}{(t)}\hat{a}_l^{\dagger}{(t)}\hat{a}_k{(t)}+\mathcal{D}_{kl}{(t)}\hat{a}_k{(t)}\hat{a}_l{(t)}\right],\nonumber
\end{align}
{which is: (i) hermitian, by satisfying the conditions $\mathcal{A}_{kl}{(t)}=\mathcal{D}_{kl}^*{(t)}$, $\mathcal{B}_{kl}{(t)}=\mathcal{C}_{kl}^*{(t)}$ and (ii) invariant over an index change, with the conditions} $\mathcal{A}_{kl}{(t)}=\mathcal{A}_{lk}{(t)}$, $\mathcal{D}_{kl}{(t)}=\mathcal{D}_{lk}{(t)}$, $\mathcal{B}_{kl}{(t)}=\mathcal{C}_{lk}{(t)}$ and $\mathcal{B}_{lk}{(t)}=\mathcal{C}_{kl}{(t)}$.

Suppressing the notation for time dependence, the correspondent Heisenberg equation of motion for the annihilation and creation operators is therefore
\begin{align}
    \dot{\hat{a}}_j =i\left[\hat{H},\hat{a}_j\right]=i\sum_{kl}\Bigg(&\mathcal{A}_{kl}\qty[\hat{a}_k^{\dagger}\hat{a}_l^{\dagger},\hat{a}_j]+\mathcal{B}_{kl}\qty[\hat{a}_k^{\dagger}\hat{a}_l,\hat{a}_j]\nonumber\\
&+\mathcal{C}_{kl}\qty[\hat{a}_l^{\dagger}\hat{a}_k,\hat{a}_j]+\mathcal{D}_{kl}\qty[\hat{a}_k\hat{a}_l,\hat{a}_j]\Bigg)\nonumber\\
    \label{daH}=-i\sum_k\bigg[\left(\mathcal{A}_{kj}+\mathcal{A}_{jk}\right)&\hat{a}_k^{\dagger}+\left(\mathcal{B}_{jk}+\mathcal{C}_{kj}\right)\hat{a}_k\bigg]
\end{align}
and
\begin{align}
    \dot{\hat{a}}_j^{\dagger} =i\left[\hat{H},\hat{a}_j^{\dagger}\right]=i\sum_{kl}&\Bigg(\mathcal{A}_{kl}\qty[\hat{a}_k^{\dagger}\hat{a}_l^{\dagger},\hat{a}_j^{\dagger}]+\mathcal{B}_{kl}\qty[\hat{a}_k^{\dagger}\hat{a}_l,\hat{a}_j^{\dagger}]\nonumber\\
    &+\mathcal{C}_{kl}\qty[\hat{a}_l^{\dagger}\hat{a}_k,\hat{a}_j^{\dagger}]+\mathcal{D}_{kl}\qty[\hat{a}_k\hat{a}_l,\hat{a}_j^{\dagger}]\Bigg)\nonumber\\
    =\label{dadH}i\sum_k\bigg[\left(\mathcal{D}_{kj}+\mathcal{D}_{jk}\right)&\hat{a}_k+\left(\mathcal{B}_{kj}+\mathcal{C}_{jk}\right)\hat{a}_k^{\dagger}\bigg].
\end{align}

Comparing \eqref{da} with \eqref{daH} and \eqref{dad} with \eqref{dadH}, we obtain the following system
\begin{subequations}
\label{coef}
    \begin{align*}
    -i\left[\mathcal{A}_{kj}{(t)}+\mathcal{A}_{jk}{(t)}\right]&=-2i\mathcal{A}_{kj}{(t)}=B_{kj}^*{(t)}\\
    -i\left[\mathcal{C}_{kj}{(t)}+\mathcal{B}_{jk}{(t)}\right]&=-2i\mathcal{C}_{kj}{(t)}=A_{kj}{(t)}\\
    i\left[\mathcal{D}_{kj}{(t)}+\mathcal{D}_{jk}{(t)}\right]&=2i\mathcal{D}_{kj}{(t)}=B_{kj}{(t)}\\
    i\left[\mathcal{B}_{kj}{(t)}+\mathcal{C}_{jk}{(t)}\right]&=2i\mathcal{B}_{kj}{(t)}=A_{kj}^*{(t)}.
\end{align*}
\end{subequations}
{Inserting the last coefficients into Eq.~\eqref{ht}, one obtains the following expression for the} effective Hamiltonian
\begin{align}
\hat{H}_{{H}}{(t)} &=\frac{i}{2}\sum_{{j} {k}} \Bigg[A_{kj}{(t)}\hat{a}_{j}^{\dagger}{(t)}\hat{a}_{k}{(t)}+B_{kj}^*{(t)}\hat{a}_{j}^{\dagger}{(t)}\hat{a}_{k}^{\dagger}{(t)}
-\text{h.c.}\Bigg],
\end{align}
{where the subscript $H$ conveys that the operator is represented in the Heisenberg picture of quantum mechanics.}

{Moving to the Schrodinger picture, the last Hamiltonian takes the form
\begin{align}
\label{HS}
\hat{H}_S(t) &=\frac{i}{2}\sum_{{j} {k}} \Bigg[A_{kj}(t)\hat{b}_{j}^{\dagger}\hat{b}_{k}+B_{kj}^*(t)\hat{b}_{j}^{\dagger}\hat{b}_{k}^{\dagger}-\text{h.c.}
\Bigg],
\end{align}
where the Heisenberg annihilation (and creations) operator is defined as $\hat{a}_k(t)=\hat{U}_S^{\dagger}(t)\hat{b}_k\hat{U}_S(t)$, with $\hat{U}_S(t)$ being the time evolution operator generated by the Hamiltonian \eqref{HS}}.

\end{document}